\documentclass[aps,pre,twocolumn,reprint]{revtex4-1}
\usepackage{amsmath}
\usepackage{graphicx,}
\usepackage{dcolumn}
\usepackage{bm}
\usepackage{amssymb}
\usepackage{float}
\usepackage{epstopdf}
\usepackage{wasysym}
\usepackage{color}
\begin{document}	
	\author{Ahmad K. Omar}
	\email{aomar@berkeley.edu}
	\author{Zhen-Gang Wang}
	\email{zgw@caltech.edu}
	\author{John F. Brady}
	\email{jfbrady@caltech.edu}
	\affiliation{Division of Chemistry and Chemical Engineering, California Institute of Technology, Pasadena, California 91125, USA}
	
	\title{Microscopic Origins of the Swim Pressure and the Anomalous Surface Tension of Active Matter}
	\date{Submitted June 6, 2019; to appear in Physical Review E}
	\begin{abstract}
	The unique pressure exerted by active particles -- the ``swim" pressure -- has proven to be a useful quantity in explaining many of the seemingly confounding behaviors of active particles. However, its use has also resulted in some puzzling findings including an \textit{extremely negative} surface tension between phase separated active particles. Here, we demonstrate that this contradiction stems from the fact that the swim pressure \textit{is not a true pressure}. At a boundary or interface, the reduction in particle swimming generates a net active force density -- an entirely \textit{self-generated body force}. The pressure at the boundary, which was previously identified as the swim pressure, is in fact an elevated (relative to the bulk) value of the \textit{traditional particle pressure} that is generated by this interfacial force density. Recognizing this unique mechanism for stress generation allows us to define a much more physically plausible surface tension. We clarify the utility of the swim pressure as an ``equivalent pressure" (analogous to those defined from electrostatic and gravitational body forces) and the conditions in which this concept can be appropriately applied.
	\end{abstract}

\maketitle

\section{Introduction} 
While the development of a formal nonequilibrium statistical description of active particles remains an exciting and ongoing challenge~\cite{Fodor2016, Speck2016b, Marconi2017, Puglisi2017, Nardini2017, Mandal2017, Grandpre2018, Shankar2018, Dabelow2019}, \textit{mechanical descriptions} have proven to be a powerful tool in describing many of the seemingly confounding behaviors of active particles. Work, pressure and tension are well-defined mechanical concepts and can thus be computed for materials arbitrarily far from equilibrium. In recent years, the pressure of active matter~\cite{Mallory2014a, Fily2014, Takatori2014, Solon2015a, Solon2015c, Epstein2019} has aided in the description of many phenomena including instabilities exhibited by expanding bacterial droplets~\cite{Sokolov2018}, the dynamics of gels~\cite{Szakasits2017, Omar2019} and membranes~\cite{Paoluzzi2016} embedded with active particles, and even the phase behavior of living systems~\cite{Sinhuber2017}. Among the phenomena that active pressure has successfully described is the stability limit~\cite{Takatori2015, Solon2018, Paliwal2018, Solon2018} (the spinodal) of \textit{purely repulsive} active particles which are observed to separate into ``liquid-" and ``gas-like" regions, commonly referred to as motility-induced phase separation~\cite{Fily2012, Cates2015}. Yet upon using this same active pressure to compute a surface tension (cf.,~eq.~\eqref{eq:tension}) between the coexisting phases, one alarmingly finds that it is \textit{extremely negative} despite the presence of a stable (e.g.,~a tendency for the system to reduce the interfacial area as shown in Fig.~\ref{fig:nmQinterface}A) interface~\cite{Bialke2015, Patch2018}.

In this Article, we reveal that the reported anomalous surface tension points to a larger issue in the mechanics of active matter: the swim pressure~\cite{Takatori2014, Fily2014, Solon2015a} -- argued to be the nonequilibrium generalization of the equilibrium Brownian osmotic pressure -- is, in fact, \textit{not a true pressure}. By this we mean it is not a point-wise defined surface force (or stress), the relevant forces in mechanically defining the interfacial tension~\cite{Kirkwood1949}. If not a true pressure or surface force, why do swimmers exert a higher pressure on boundaries relative to passive particles? Here, we demonstrate that the enhanced pressure exerted by active particles originates from a local \textit{self-generated} active force density that arises from the active dynamics and the reduction of swimming \textit{at a boundary} (e.g.,~at a hard wall or even a gas-liquid interface). What is referred to as the swim pressure is in actuality an elevated (relative to the bulk) value of the traditional sources of pressure. The localized active force density acts as a \textit{body force} and \textit{balances a pressure difference between bulk and the boundary}. In revealing the microscopic origins of the swim pressure, we clarify its applicability and, in the process, recover a more physically plausible surface tension. 

\section{Stress Generation in Active Matter}
We begin by discussing the enabling concepts behind the swim pressure (or stress) idea. Consider a simple model for an overdamped active particle: each particle exerts a constant self-propulsive force $\bm{F^{\text{swim}}} = \zeta U_0 \bm{q}$ in a direction $\bm{q}$ in order to move at a speed $U_0$ in a medium of resistance $\zeta$. The particle orientation $\bm{q}$ undergoes random reorientation events that result in a characteristic reorientation time $\tau_R$ and run length (the distance a particle travels before reorienting) of $U_0\tau_R$. On timescales longer than $\tau_R$, these dynamics give rise to a diffusivity $D^{\text{swim}}=U_0^2\tau_R/6$ (in 3D~\cite{Berg1993}) which can be entirely athermal in origin. This swim diffusivity results in a dilute suspension of active particles with number density $n_0$ exerting a single-body \textit{diffusive} pressure on a boundary $\Pi^{\text{swim}} = n_0\zeta U_0^2\tau_R/6 = n_0\zeta D^{\text{swim}}$~\cite{Fily2014, Takatori2014, Solon2015a}. This diffusive pressure can be thought of as the nonequilibrium extension of the thermal osmotic pressure exerted by equilibrium Brownian colloids $\Pi^{B} = n_0k_BT = n_0 \zeta D_T$ (where $k_BT$ is thermal energy and $D_T$ is the Brownian diffusivity). By analogy to thermal systems, one can define an active energy scale $k_sT_s \equiv \zeta D^{\text{swim}} =  \zeta U_0^2\tau_R/6$ such that $\Pi^{\text{swim}} = n_0k_sT_s$~\cite{Takatori2015}.

Unlike the diffusive pressure of thermal Brownian colloids (the $nk_BT$ contribution to the total pressure), the swim pressure; (1) need not be isotropic (and is therefore properly a swim stress $\bm{\sigma^{\text{swim}}}$ with $\Pi^{\text{swim}} = -\text{tr}(\bm{\sigma^{\text{swim}}})/3$) as the direction of swimming could be biased (e.g.,~by an applied orienting field~\cite{Takatori2014a}); and (2) explicitly depends on the volume fraction $\phi$ of active particles. The latter effect is a consequence of interparticle interactions impeding a particle's ability to swim, reducing the actual swimming velocity (and thus, the run length and swim pressure) from the intrinsic swim speed $U_0$ with increasing particle concentration. We can include this effect as well as the influence of anisotropic swimming in the general expression for the local swim stress~\cite{Solon2015a, Takatori2014a} for particles interacting with isotropic conservative interactions (particle orientations $\bm{q}$ are independent): 
\begin{equation}
\label{eq:impulse}
\bm{\sigma^{\text{swim}}} = -\frac{\zeta U_0 U\tau_R}{2}\left[\bm{Q} + n\bm{I}/3\right],
\end{equation}
where $U$ is the magnitude of the particle velocity in the direction of swimming, $\bm{Q} = \int P(\bm{x}, \bm{q}) (\bm{qq} - \bm{I}/3) d\bm{q}$ is the traceless nematic order ($\bm{0}$ for an isotropic system), $n = \int P(\bm{x}, \bm{q})d\bm{q}$ is the local number density, $P(\bm{x}, \bm{q})$ is the probability density of an active particle having position $\bm{x}$ and orientation $\bm{q}$, and $\bm{I}$ is the identity tensor.

\begin{figure}
	\centering
	\includegraphics[width=0.48\textwidth,keepaspectratio,clip]{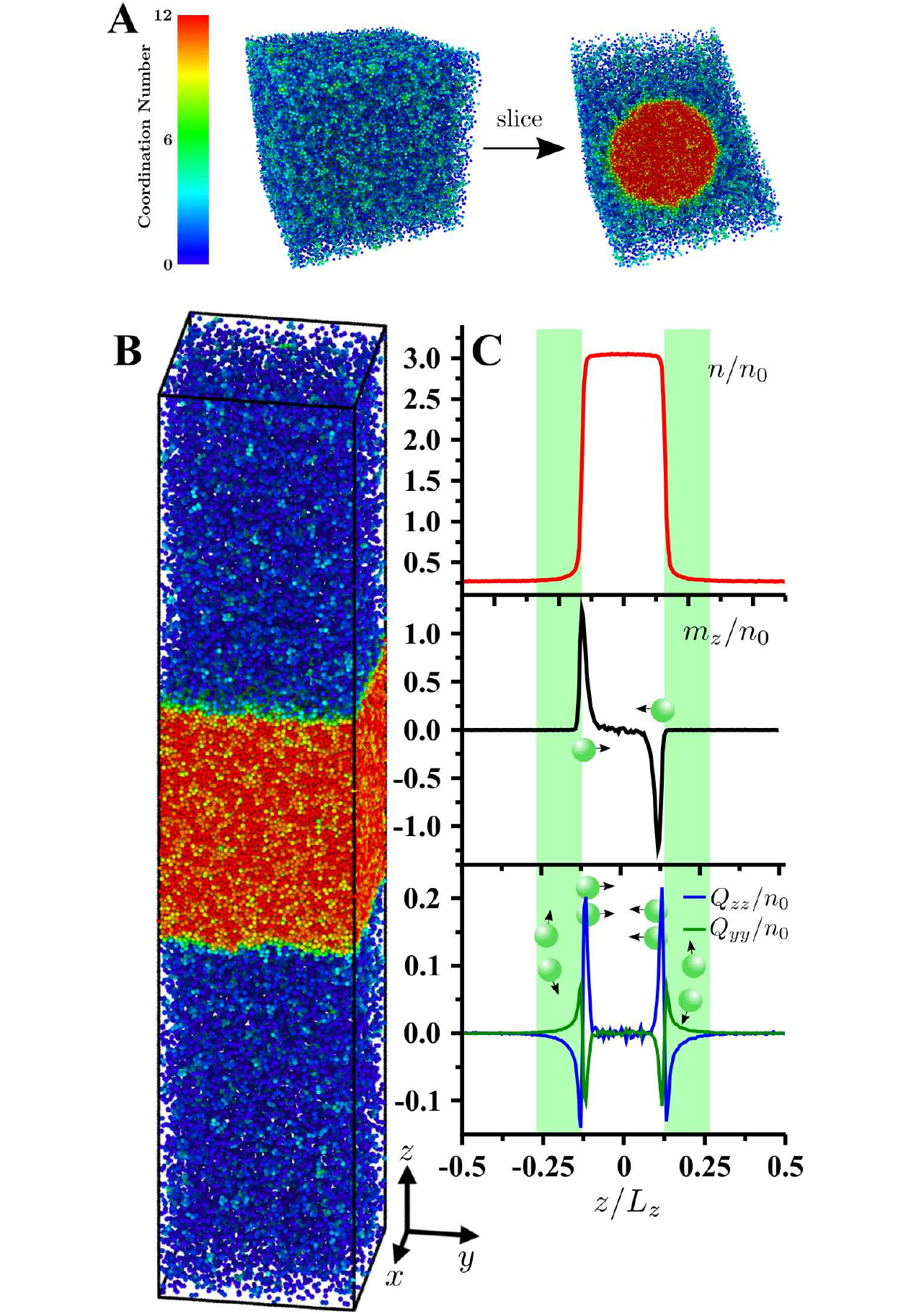}
	\caption{{\protect\small{(A) A spherical active ``liquid'' droplet with a system total of $108000$ active particles (volume fraction of $\phi = 4\pi a^3n_0/3 = 0.14$ where $n_0$ is the number density) with $\text{Pe}_R = 0.0025$. (B) A characteristic simulation snapshot for $\text{Pe}_R = 0.0025$ and $\phi=0.15$ with $148716$ particles and an asymmetric box with dimensions $L_z = 5L_x = 5L_y$. (C) The accompanying number density, polar order and nematic field profiles along the long axis ($z$) of the simulation cell. The data is translated such that the dense phase is centered along the long axis. The shaded regions are responsible for the previously reported negative surface tension ($Q_{yy} > 0$). Cartoon insets illustrate representative particle orientations.}}}
	\label{fig:nmQinterface}
\end{figure}

The reduction in swim pressure with concentration occurs for large run lengths ($U_0\tau_R \gg a$ or $\text{Pe}_R \equiv a/U_0\tau_R \ll 1$ where $a$ is the particle radius) and can lead the total pressure or the ``active pressure'' (the sum of the swim pressure and any other sources of pressure, such as interparticle interactions) to become nonmonotonic. This mechanical instability manifests through the phase separation of active particles. Figure~\ref{fig:nmQinterface}A illustrates a phase separated active matter simulation~\cite{Anderson2008,Glaser2015} for highly persistent ($\text{Pe}_R = 0.0025$), overdamped and non-Brownian ($D_T=0$) active particles interacting with a steeply repulsive WCA~\cite{Weeks1971} potential ($\text{Pe}_S \equiv \zeta U_0 a/\epsilon = 0.01$ where the Lennard-Jones diameter is taken to be $2a$ and $\epsilon$ is the Lennard-Jones energy). Full simulation details are provided in Appendix~\ref{Appendix A}. The active dynamics are fully encapsulated in $\text{Pe}_R$ and $\text{Pe}_S$, the latter of which will be held constant throughout this Article. One immediately appreciates that the liquid region forms a stable spherical domain, tending to minimize the surface area. 

While the surface tension cannot be defined thermodynamically as the excess free energy for this driven system, one can define it \textit{mechanically}~\cite{Kirkwood1949} as the ``minimum'' work required to create a differential area (at fixed volume) of interface in a planar (slab) geometry, resulting in: 
\begin{equation}
\label{eq:tension}
\gamma = -\int_{-\infty}^{+\infty} [\sigma_{zz} - \sigma_{yy}] dz,
\end{equation}
where $\sigma_{ij}$ are the components of the appropriate stress tensor $\bm{\sigma}$ and $z$ is the direction normal to the interface~\footnote{$\sigma_{xx}$ could be substituted for $\sigma_{yy}$ without loss of generality due to the isotropy of the interface in the tangential directions. For finite-sized simulation with periodic boundaries, eq.~\eqref{eq:tension} must be divided by two as there are two interfaces (see Fig.~\ref{fig:nmQinterface}) and the integration limits are now the box size.}. Upon defining the stress tensor as the sum of the swim stress and the traditional sources of particle stress $\bm{\sigma^{P}}$ (arising from interparticle interactions for our system) -- we refer to this sum as the active stress $\bm{\sigma^{\text{act}}}$ -- eq.~\eqref{eq:tension} results in a surface tension that is \textit{extremely negative} $\gamma \sim O(-nk_sT_sa)$~\cite{Bialke2015, Patch2018}, in striking contrast to our physical intuition that a mechanically stable interface must have a positive surface tension. 

In an attractive colloidal or molecular fluid, there is an excess of tangential stress (i.e.,~$\sigma_{yy} > \sigma_{zz}$ and $\gamma > 0$) within the interface. In contrast, Bialk{\'e} et al.~\cite{Bialke2015} observed that within the low density region of the interface where $U \approx U_0$ (see the shaded regions in Fig.~\ref{fig:nmQinterface}C), the particles are aligned tangential to the interface~\cite{Lee2017}, generating a strongly anisotropic local swim stress ($|\sigma^{\text{swim}}_{yy}| \gg |\sigma^{\text{swim}}_{zz}|$ where both stresses are negative) and a negative surface tension.

The problem is that the active interface cannot simply be described by the density and nematic order: an unavoidable feature of the interface is that the particles, on average, point towards the liquid phase as particles pointing towards the gas are free to escape. This polarization of active particles can be quantified through the polar order defined as $\bm{m} = \int P(\bm{x}, \bm{q})\bm{q}d\bm{q}$ as is shown in Fig.~\ref{fig:nmQinterface}C. This polarization of the particles results in volume elements within the interface having a \textit{swim force density} $\zeta U_0 \bm{m}$. 

\begin{figure}
	\centering
	\includegraphics[width=0.48\textwidth,keepaspectratio,clip]{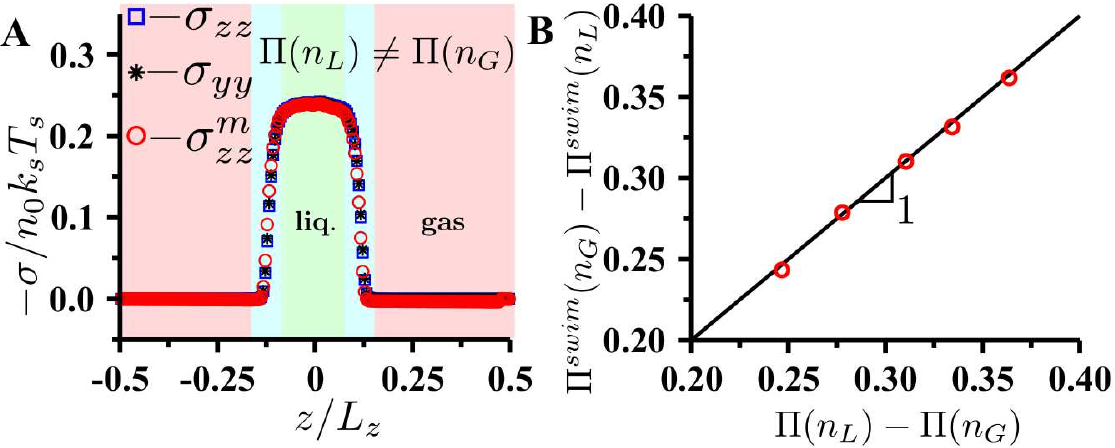}
	\caption{{\protect\small{(A) The measured components of the interparticle stress and the stress predicted through the integral of the swim force density for the identical system in Fig.~\ref{fig:nmQinterface}C. (B) The difference in the swim pressure between the gas and liquid phases as a function of the interparticle pressure difference between the liquid and gas phases. Each point represents a distinct value of $\text{Pe}_R$. Pressures are normalized by $n_0k_sT_s$.}}}
	\label{fig:stress}
\end{figure}

It is important to recognize that while this interfacial force density emerges naturally -- it is \textit{internally} generated -- its role will be no different than an externally applied body force (e.g.,~gravity). In the absence of particle flow, acceleration or any applied external forces, a simple point-wise momentum balance on the active particles \textit{must} result in: 
\begin{equation}
\label{eq:balance}
\nabla \cdot \bm{\sigma} + \zeta U_0 \bm{m} = \bm{0},
\end{equation}
where $\bm{\sigma}$ is the stress that must balance the force density created by the polarization of the active particles. From eq.~\eqref{eq:balance} we can immediately recognize that there will be a rapid stress variation across the interface due to the localized swim force density: \textit{the liquid and gas phases have different pressures}. We further examine this breakdown of the commonly presumed coexistence criterion of pressure equality by integrating the swim force density profile found in simulation to obtain the predicted stress (or pressure) profile  ($\sigma^m_{zz}$) up to an additive constant. As shown Fig.~\ref{fig:stress}A, the liquid and gas phase pressures are indeed strikingly disparate and the predicted stress profile precisely matches the interparticle stress ($\bm{\sigma^P}$): $\bm{\sigma}$ in eq.~\eqref{eq:balance} \textit{does not include the swim stress} and is simply $\bm{\sigma^P}$, which, for our system, is simply the stress arising from conservative interparticle interactions. We can mechanically describe the system \textit{without any notion of swim pressure.}

\begin{figure}
	\centering
	\includegraphics[width=0.48\textwidth,keepaspectratio,clip]{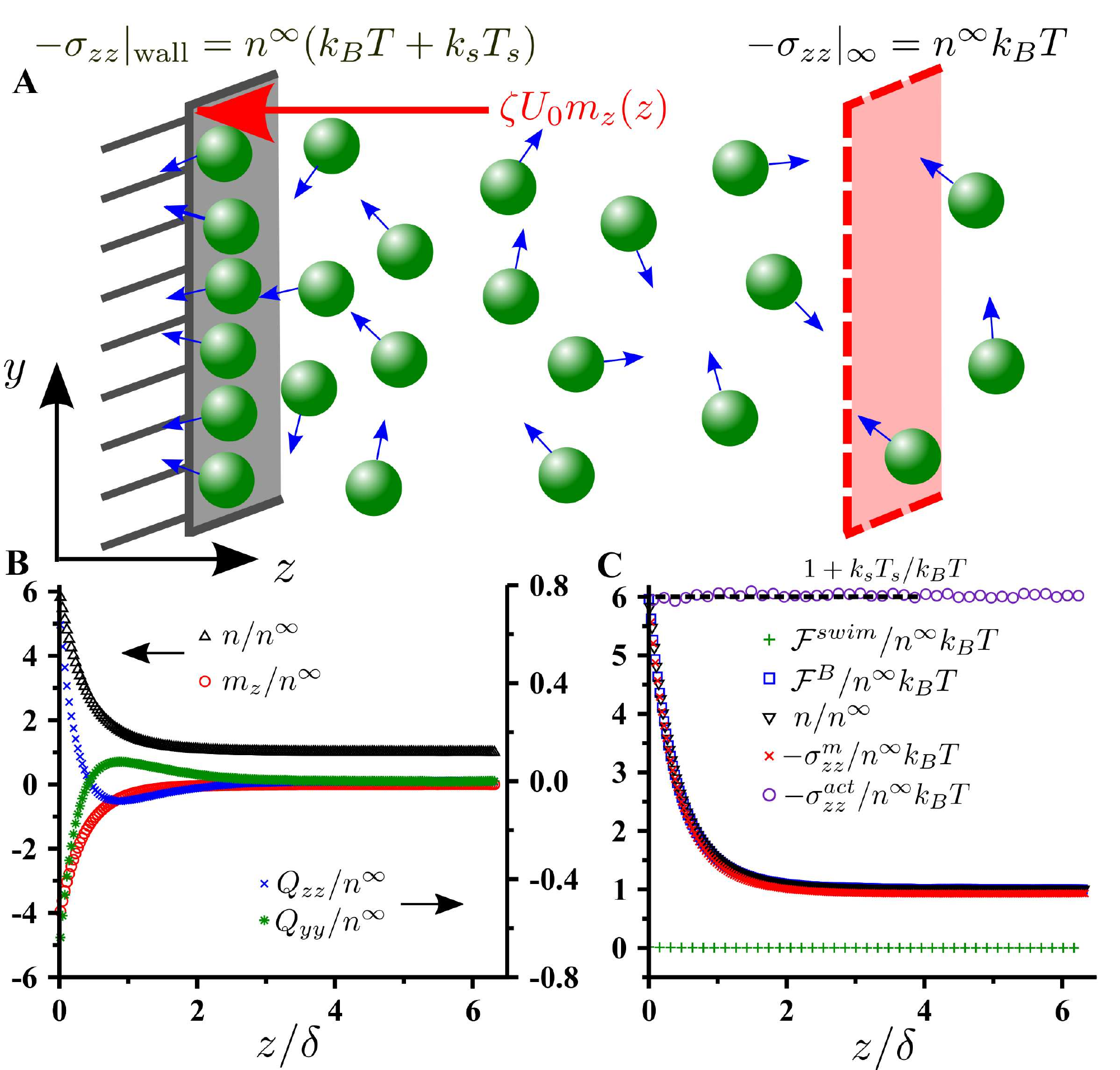}
	\caption{{\protect\small{(A) Schematic system of active Brownian particles near a hard wall. For $k_sT_s/k_BT = 5$ and a 2D system (B) the number density, polar order and nematic fields and the (C) force flux, number density and local stress profiles.}}}
	\label{fig:wall}
\end{figure}

Understanding the above finding requires revisiting the microscopic origins of the swim stress. Consider a simple 2D system of ideal (noninteracting) active Brownian particles (ABPs) in the presence of an impenetrable wall with a normal in the $+z$-direction (see Fig.~\ref{fig:wall}A). The measured wall pressure is $n^{\infty}(k_BT + k_sT_s)$, and in the absence of flow, acceleration and \textit{externally} applied body forces, previously led to the conclusion that active particles exert a mechanical swim pressure that is spatially homogeneous. However, the active particles accumulate on and \textit{orient} towards the boundary (see Fig.~\ref{fig:wall}B) with a thickness proportional to a microscopic length scale $\delta=\sqrt{D_T\tau_R}$~\cite{Yan2015}. From our previous discussion, we now recognize that the presence of a swim force density $\zeta U_0 m_z$ within the boundary layer \textit{must be} considered in the momentum balance. This suggests that, in contrast to most studies (notwithstanding~\cite{Speck2016}), the \textit{stress is not spatially constant}. Figure~\ref{fig:wall}C reveals that the stress profile found by integrating $\zeta U_0 m_z$ is precisely the anticipated Brownian osmotic stress ($-\sigma_{zz}^{m} = n(z)k_BT$). Just as before, $\bm{\sigma}$ in eq.~\eqref{eq:balance} is simply the traditional sources of stress $\bm{\sigma^P}$ and \textit{does not include the swim stress}. We have explicitly found (using a procedure~\cite{Todd1995} described in the Supplemental Material~\footnote{See the Supplemental Material at [URL will be inserted by publisher] for a discussion (which includes Refs.~\cite{Takatori2014,Todd1995}) of the local stress tensor and method-of-planes method as well as simulation movies }.) that the local stress generated by the Brownian force $-\mathcal{F}^B$ is precisely $-n(z)k_BT$ while that generated by the swim force $-\mathcal{F}^{\text{swim}}$ is \textit{negligible}. We further note that for ideal ABPs the absence of the swim stress can be rigorously shown to be true as the flux of density $n$ is zero everywhere $\bm{j_{n}} = -D_T\nabla n + U_0\bm{m} = 0$ which is equivalent to eq.~\eqref{eq:balance} with $\bm{\sigma} = \bm{\sigma^P} = -nk_BT \bm{I}$.

Further, consider inserting a wall into the bulk region of the active particles as depicted in Fig.~\ref{fig:wall}A. One would \textit{instantaneously} measure a stress of $-\sigma_{zz} = -\sigma^P_{zz} = n^{\infty}k_BT$ as it is only after a time $\tau_R$ that the accumulation boundary layer forms and the resulting swim force density raises the pressure $-\sigma^P_{zz}$ at the wall to be $n^{\infty}(k_BT + k_sT_s)$ (via increasing the density). It was previously shown that the details of the particle-wall interaction can alter the measured pressure exerted on the boundary~\cite{Solon2015c}. This observation lead to the conclusion that active matter does not generally admit an equation-of-state as the pressure in bulk and at the boundary may differ if the boundary exerts torques on the active particles. Our findings illustrate that even in the absence of such particle-wall interactions, there is \textit{always a pressure difference between the bulk and the boundary} and the self-generated swim force density balances this difference.

\section{Origins and Applicability of the Swim Stress}
Why is it that the swim force density which balances a pressure difference between the wall and boundary is \textit{precisely} the swim pressure?  To address this, we turn to the steady-state conservation equation for the polar order field which can readily be derived from the full Smoluchowski equation~\cite{Saintillan2015, Yan2015, Fily2017, Solon2018a, Paliwal2018} as:
\begin{equation}
\label{eq:mequation}
-\nabla \cdot \bm{j_m} - \frac{2}{\tau_R} \bm{m} + \bm{\Gamma} = \bm{0},
\end{equation}
where $\Gamma$ represents any externally applied sink or sources of polar order (including torque-exerting boundaries~\cite{Solon2015c, Yan2015a, Fily2017}) and $-2\bm{m}/\tau_R$ is a natural sink that arises due to rotary diffusion of the active particles. The flux of polar order is $\bm{j_m} = U[\bm{Q} + n\bm{I}/3] - D_T\nabla \bm{m}$. Substituting the above expression into eq.~\eqref{eq:balance} gives:
\begin{equation}
\nabla\cdot\bm{\sigma^P} + \frac{1}{2}\zeta U_0\tau_R \left( -\nabla \cdot \bm{j_m} + \bm{\Gamma} \right) = \bm{0}. 
\end{equation} 
Thus, near a planar no flux boundary the pressure difference between the boundary and bulk \textit{must be} $-\sigma^{P}_{zz}|_{\text{wall}} + \sigma^{P}_{zz}|_{\infty} = \frac{1}{2}\zeta U_0 \tau_R (-j_{m,zz}|_{\text{wall}}+j_{m,zz}|_{\infty})=n^{\infty}k_sT_s$. We can also recognize that many of the interesting dependencies of the force on a boundary exerted by active matter can now all be understood within this perspective. The dependence of this force on the boundary curvature~\cite{Smallenburg2015, Yan2015, Yan2018}, particle-boundary interactions~\cite{Solon2015c}, and other details that \textit{would not} affect the pressure of passive matter naturally follows from the sensitivity of the active force density (polar order) to these details and the coupling of polar order and stress through eq.~\eqref{eq:balance}.

How can we understand the absence of the swim stress from the above discussion yet its success in describing a host of behaviors? Using eq.~\eqref{eq:mequation}, we can express the momentum balance eq.~\eqref{eq:simplifiedmomentum}~\footnote{Equation~\eqref{eq:simplifiedmomentum} assumes that the swim speed and reorientation time of the particles are spatially constant and density independent as otherwise an additional body force proportional to $\nabla U_0(\bm{x}) \tau_R(\bm{x})$ would arise in the momentum balance.} as:
\begin{equation}
\label{eq:simplifiedmomentum}
\nabla \cdot \bm{\sigma^{\text{act}}} + \frac{1}{2} \tau_R \zeta U_0 \bm{\Gamma} = \bm{0},
\end{equation}
where $\bm{\sigma^{\text{act}}} = \bm{\sigma^P} - \frac{1}{2} \tau_R \zeta U_0\bm{j_m} = \bm{\sigma^P} + \bm{\sigma^{\text{swim}}} + \frac{1}{2} \tau_R \zeta U_0 D_T\nabla \bm{m}$. Equation~\eqref{eq:simplifiedmomentum} is the frequently used continuum momentum balance~\cite{Yan2015a, Fily2017} but it is crucial to appreciate that $\bm{\sigma^{\text{act}}}$ is \textit{no longer the system stress as it contains elements from the original body force} $\zeta U_0 \bm{m}$ (those that could be expressed as a divergence of a tensor), recast as $\bm{\sigma^{\text{swim}}}$. The true stress remains $\bm{\sigma^P}$. This is analogous to the pressure field $p$ of a static liquid of density $\rho$ subject to a gravitational field $g$ (acting in the $-z$-direction). The momentum balance for this system $\nabla p + \rho \bm{g} = \bm{0}$ is often expressed as $\nabla \mathcal{P} = \bm{0}$ where $\mathcal{P} = p + \rho g z$ is often referred to as an ``equivalent" pressure. One would obviously not conclude that the hydrostatic pressure is independent of the depth simply because $\mathcal{P}$ is a constant -- the true pressure is $p$ just as the true stress of active matter is encapsulated in $\bm{\sigma^{P}}$, with the swim stress playing a similar role as the gravitational potential $\rho g z$~\cite{Paliwal2018}. A similar analogy can be made between the swim stress and the Maxwell stress in electrostatics, which represents the body force acting on charge density from an electric field~\cite{Woodson1968}. We further note that in the more generalized momentum balance which includes the transient terms in the conservation equations (e.g.,~eqs.~\eqref{eq:balance} and \eqref{eq:mequation}) derived by Epstein et al.~\cite{Epstein2019}, one cannot readily absorb the swim stress into the true stress tensor to define the active stress. The active stress is therefore only \textit{rigorously} applicable in the steady state or quasi-steady state (e.g.,~slowly relaxing polar order field).

\section{Surface Tension of Phase Separated Active Particles}
With the origin of the swim stress now more clearly established, we can begin to decipher the utility as well as the potential pitfalls of invoking it by returning to the context of active phase separation. In the absence of external sources/sinks of polar order (i.e.,~no net torques anywhere in the system $\bm{\Gamma} = \bm{0}$), invoking the swim stress and eq.~\eqref{eq:simplifiedmomentum} implies a spatially constant active stress (confirmed for the wall situation in Fig.~\ref{fig:wall}C) and thus restores the convenient coexistence criterion of equal mechanical pressures between the liquid and gas phases. Indeed, the difference in interaction pressure between the two phases is equal and opposite to the difference in swim pressures (see Fig.~\ref{fig:stress}B). Simply knowing that the particles can rotate freely allows one to invoke the active stress perspective and bypass solving for the swim force density within microscopic boundary layers \textit{so long as one is not looking to define the stress at a point in space}.

\begin{figure}
	\centering
	\includegraphics[width=0.48\textwidth,keepaspectratio,clip]{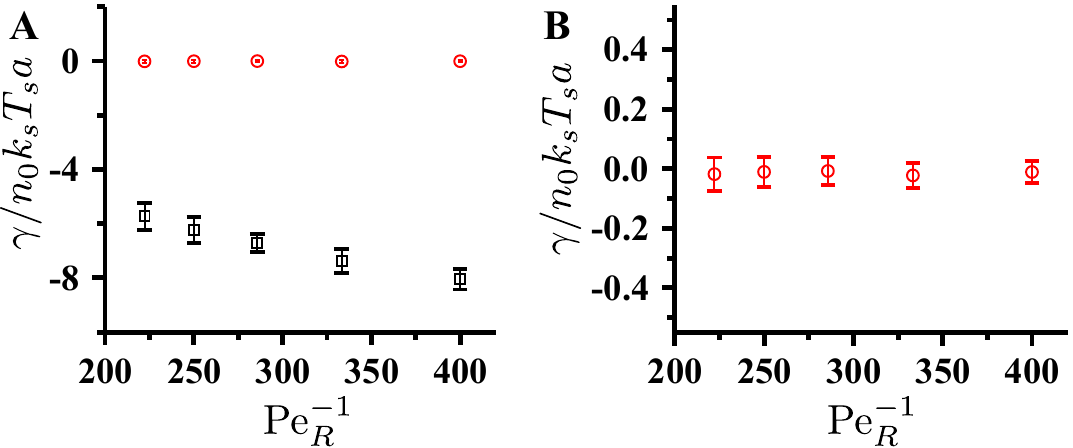}
	\caption{{\protect\small{(A) The surface tension of active particles obtained through use of the true stress ($\ocircle$) of active particles (see (B) for a magnified view) in comparison to that obtained using the active stress ($\square$).}}}
	\label{fig:surfacetension}
\end{figure}

Despite its utility as a phase coexistence criterion, using the active stress to compute the surface tension results in the extremely negative interfacial tension (see Fig.~\ref{fig:surfacetension}A) that strongly contrasts with our physical observations. We now recognize that this is because the surface tension requires use of the true stress locally exerted by the particles $\bm{\sigma^P}$. By using the correct stress in eq.~\eqref{eq:tension} (which remains valid in the presence of a body force as shown in Appendix~\ref{Appendix B}), we find that the surface tension is almost negligible and displays little dependence on the level of particle activity. One can appreciate the smallness of $\gamma$ through the isotropy in the stress (Fig.~\ref{fig:stress}).

That the surface tension is vanishingly small (rather than significantly negative) is reassuring, but might suggest that the active interface should be quite volatile. We note that relating the interfacial height fluctuations of driven systems to surface tension using standard capillary wave theory (CWT) is problematic as the theory is formulated using equilibrium statistical physics. Studies on the interface of driven systems that have used CWT explicitly included thermal noise in their systems and implicitly made the ansatz that thermal fluctuations dominate over nonequilibrium effects~\cite{Derks2006, Paliwal2017, Patch2018, Junco2019} (which clearly is not applicable for our athermal system) or have substituted the ``housekeeping work" in place of the thermal energy~\cite{Bialke2015, Speck2016b}. In addition to characterizing the athermal source of fluctuations, the influence of numerous mechanical factors (beyond the intrinsic surface tension measured in this work) must be explored including the potential  bending stiffness~\cite{Patch2018} of the interface and understanding if the swim force density plays a similar role as traditional external body forces (e.g.,~gravity~\cite{Buff1965}) in suppressing interfacial height fluctuations.

Our findings have implications that extend beyond resolving the controversy of a deeply negative surface tension. \textit{All components of} $\bm{\sigma^{\text{swim}}}$ are not the true stresses exerted by particles in bulk but \textit{might be} relevant at a boundary. The mechanism by which the off-diagonal components of $\bm{\sigma^{\text{swim}}}$ (e.g.,~shear swim stresses~\cite{Takatori2017, Saintillan2018}) are transmitted to a boundary is not immediately obvious and merits further investigation. Even in the absence of a torque-inducing wall~\cite{Solon2015c} (see eq.~\eqref{eq:simplifiedmomentum}), measuring the force on a boundary in ``wet" active matter systems requires recognizing that the spatially constant sum of the active particle $\Pi^P$ and fluid $p_f$ pressures will have a value far from the boundary (and thus, everywhere) \textit{which does not include the swim pressure.} Only by isolating the value of $\Pi^P$ at the boundary can the swim pressure be directly isolated.

\begin{acknowledgments}
A.K.O. acknowledges support by the National Science Foundation Graduate Research Fellowship under Grant No. DGE-1144469 and an HHMI Gilliam Fellowship. J.F.B. acknowledges support by the National Science Foundation under Grant No. CBET-1803662. 
\end{acknowledgments}

\appendix
\section{Simulation and Calculation Details}
\label{Appendix A}
\subsection{Interacting, Athermal Active Particles}
In all simulations except for those shown in Fig.~3 (the details for those simulations are provided below), the motion of particle $i$ is governed by the overdamped Langevin equation $\bm{F_i^{\text{swim}}} + \sum_{j\ne i}^{}\bm{F_{ij}^{P}} - \zeta \bm{U_i} = \bm{0}$ where $\bm{F^{\text{swim}}_i} = \zeta U_0 \bm{q_i}$ is the swim force, $\bm{F^P_{ij}}$ is interparticle force from particle $j$, and $\bm{U_i}$ is the instantaneous particle velocity. The orientation dynamics also follow an overdamped Langevin equation $\bm{L_i^{R}} - \zeta_R\bm{\Omega_i} = \bm{0}$ where $\bm{\Omega_i}$ is the angular velocity of $\bm{q_i}$, $\bm{L_i^{R}}$ is the random reorientation torque and $\zeta_R$ is the rotational drag. Note that the rotational drag has no dynamical consequences as we can rewrite the angular equation-of-motion as $\bm{\tilde{L}_i^R} - \bm{\Omega_i} = \bm{0}$ with a redefined torque $\bm{\tilde{L}^R_i}$ which has white noise statistics $\overline{\bm{\tilde{L}_i^R}} = \bm{0}$ and $\overline{\bm{\tilde{L}_i^R}(t)\bm{\tilde{L}_j^R}(0)} = 2\delta(t)\delta_{ij}\bm{I}/\tau_R $ where $\delta(t)$ and $\delta_{ij}$ are Dirac and Kroneker deltas, respectively. These orientation dynamics give rise to a rotational diffusivity $\tau_R^{-1}$ that need not be thermal in origin. We emphasize that these equations of motion are \textit{entirely athermal} as we do not include (thermal) Brownian motion.

The interparticle force is derived from a steeply repulsive WCA potential~\cite{Weeks1971} with an interaction energy $\epsilon$ and a Lennard-Jones diameter of $2a$. Dimensional analysis of the equations of motion reveals that the dynamics are completely described by the reorientation  P\'{e}clet number $\text{Pe}_R \equiv a/U_0\tau_R$ and a swim  P\'{e}clet number $\text{Pe}_S \equiv \zeta U_0 a /\epsilon$. The phase behavior of hardsphere active particles is entirely controlled by the run length of the particles ($\text{Pe}_{R}$)~\cite{Takatori2015}. However, for finite particle softness there can additionally be a swim force ($\text{Pe}_{S}$) dependence and we therefore hold $\text{Pe}_S = 0.01$ fixed as a control for all of our simulations.

For the isotropic simulation shown in Fig.~\ref{fig:nmQinterface}A, the particles were initially placed in an FCC packing with a lattice constant of $3.47a$. The resulting crystal is centered within the simulation box and does not fill the entire box. This initial configuration biases the system towards rapidly forming a single liquid-droplet rather than multiple liquid domains scattered throughout the box. The latter situation would require longer simulation times to allow the isolated liquid domains to coalesce into a single drop. The simulation was run for a duration of $13000 a/U_0$. For the slab geometries, the particles were initially placed in a space-spanning FCC packing with a reduced initial box size $L_{z0}$ in the $z$-direction and a final box size of $L_{z}=2.66L_{z0}$. The box is symmetrically elongated about the $z$-axis at a speed of $\approx 0.25 U_0$ until a length of $L_z$ is achieved. This procedure again biases the formation of a single liquid domain. Upon reaching the final box size, the system is evolved for $\approx 9000 a/U_0$. The data displayed in the figures in the main text are the block average of data collected during the final $2000 a/U_0$ of the simulations and error bars represent the standard deviation of the data sampled over this time. All simulations were performed using the GPU-enabled HOOMD-blue molecular dynamics package~\cite{Anderson2008,Glaser2015}.

\begin{figure}
	\centering
	\includegraphics[width=0.35\textwidth,keepaspectratio,clip]{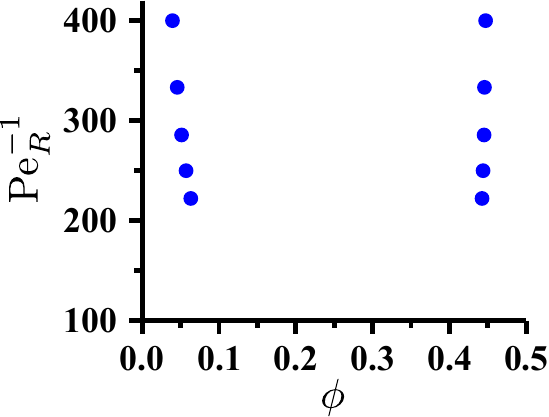}
	\caption{{\protect\small{Region of the coexistence curve explored in this work (obtained via simulation).}}}
	\label{fig:coexist}
\end{figure}

The interaction stress $\bm{\sigma}^P$ was computed using the standard virial approach with $\bm{\sigma}^P = -n \langle \bm{x_{ij}}\bm{F_{ij}^{P}} \rangle$ where $\bm{x_{ij}}$ is the distance between particles $i$ and $j$, $n$ is the local number density of the system, and the brackets denote an ensemble average over all particle pairs. The local swim stress is computed using eq.~\eqref{eq:impulse}. The local number density, polar order, nematic order and stress profiles are found by dividing the slab geometry into bins of thickness $\delta z \approx 2.4 a$ in the $z$-direction and averaging over the particles within each bin. The swim pressure difference between the liquid and gas phases shown in Fig.~\ref{fig:stress}B were found using the local value of the swim stress in the two phases for various values of $\text{Pe}_R$. The region of the coexistence curve examined is shown in Fig.~\ref{fig:coexist}. 

\subsection{Noninteracting Active Brownian Particles}

The system simulated in Fig.~\ref{fig:wall} consisted of noninteracting $\bm{F_{ij}} = \bm{0}$ active Brownian particles (ABPs) with an equation of motion $\bm{F_i^{\text{swim}}} + \bm{F^B} + \bm{F^{\text{wall}}} - \zeta \bm{U_i} = \bm{0}$ where we have now introduced a stochastic Brownian force with white noise statistics $\overline{\bm{F^B}} = \bm{0}$ and $\overline{\bm{F_i^B}(t)\bm{F_j^B}(0)} = 2k_BT\zeta \delta(t)\delta_{ij}\bm{I}$. The presence of an impenetrable wall is reflected in the force the wall must exert on a particle to prevent it from penetrating the boundary. The reorientation dynamics are identical to those described for the interacting system described above. We choose to simulate a system with modest activity ($k_sT_s/k_BT = 5$) such that we can easily resolve the boundary layer which becomes increasingly thin with increasing activity~\cite{Yan2015}. 

\section{Surface Tension Definition}
\label{Appendix B}
\begin{figure}
	\centering
	\includegraphics[width=0.48\textwidth,keepaspectratio,clip]{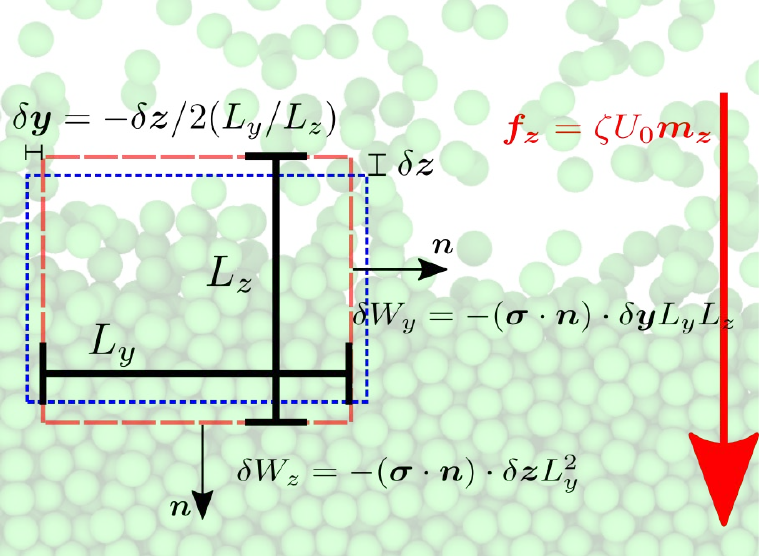}
	\caption{{\protect\small{Schematic of the interfacial mechanical balance used to define the surface tension. The dashed red box represents a 2D projection of the original control volume and the dotted blue line represents the isochorically deformed volume.}}}
	\label{fig:surfacetensionsm}
\end{figure}
Let us revisit the mechanical definition of surface tension in order to explore if a force density within the interface alters the traditional definition (eq.~\eqref{eq:tension} in the main text). Consider a rectangular control volume within the interface, shown schematically in Fig.~\ref{fig:surfacetensionsm}. The interfacial tension is typically defined as the work required to expand the box in the tangential ($y$ and $x$) directions by a width $\delta y$ while compressing the volume in the normal direction ($z$) by a width $\delta z$ such that the total volume is conserved. The latter constraint results in $\delta y = -\delta z L_y/2L_z$ where $L_y$ and $L_z$ are shown schematically in Fig.~\ref{fig:surfacetensionsm}. We note that the $x$ and $y$ directions are equivalent.  

The work required to displace a surface of the control volume is directly proportional to the true surface stress acting on the surface of interest. The presence of a body force ($\zeta U_0 \bm{m}$) within the interface results in a normal stress $\sigma_{zz}$ variation across the interface, a feature that distinguishes an active interface from traditional equilibrium interfaces which only exhibit tangential stress variation $\sigma_{yy}$. We therefore take the limit of $L_z \to dz$ (where $dz$ is a differential length) such that now the local stresses are approximately constant across the control volume. Adding the work required to move each of the six faces of the now infinitesimal volume results in: 
\begin{equation}
\label{eq:surfacework}
\delta W = -\delta A[\sigma_{zz} - \sigma_{yy}] dz,
\end{equation}
where $\delta A = 2 L_y \delta y$ is the change in tangential surface area of the system. We integrate this expression across the normal direction to obtain the total work required to expand the interface:
\begin{equation}
\label{eq:work}
W = -\delta A \int_{-\infty}^{+\infty}[\sigma_{zz} - \sigma_{yy}]dz,
\end{equation}
where we can now invoke that the definition of the interfacial tension as $W/\delta A$ with:
\begin{equation}
\label{eq:tensionsm}
\gamma = -\int_{-\infty}^{+\infty}[\sigma_{zz} - \sigma_{yy}]dz,
\end{equation}
where we assume that only a single interface is present within the system. This is identical to the traditional mechanical definition of surface tension and highlights that the presence of a body force has no \textit{explicit} effect on the surface tension; it must be recalled, however, that $\sigma_{zz}$ now varies across the interface due to the local swim force.

\end{document}


\author{Ahmad K. Omar}
	\email{aomar@berkeley.edu}
	\author{Zhen-Gang Wang}
	\email{zgw@caltech.edu}
	\author{John F. Brady}
	\email{jfbrady@caltech.edu}
	\affiliation{Division of Chemistry and Chemical Engineering, California Institute of Technology, Pasadena, California 91125, USA}
	
	\title{Supplemental Material -- Microscopic Origins of the Swim Pressure and the Anomalous Surface Tension of Active Matter}

\maketitle

\section{Local Pressure Calculation and Discussion}

\begin{figure}
	\centering
	\includegraphics[width=0.48\textwidth,keepaspectratio,clip]{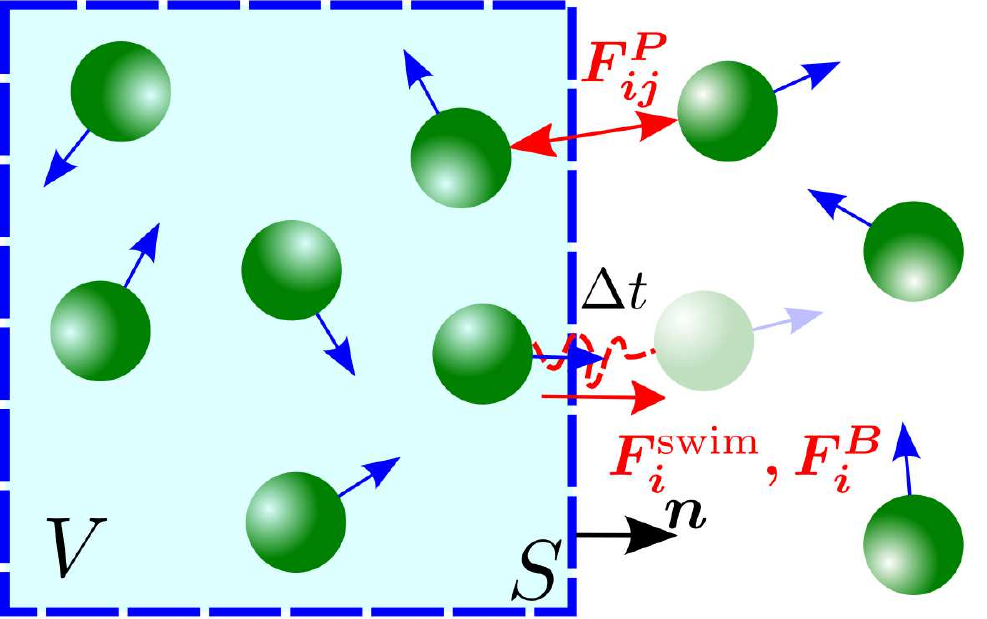}
	\caption{{\protect\small{Schematic for computing the local stress (flux of force across a plane).}}}
	\label{fig:forceflux}
\end{figure}

The local pressure -- the flux of force across the surface $S$ of a control volume $V$ (cf.~Fig.~\ref{fig:forceflux})-- exerted by ABPs was computed and shown in Fig.~3C in the main text by making the following elemental arguments. From a thermodynamic perspective, the ideal Brownian osmotic pressure of colloids $\Pi^B = nk_BT$ is the result of the translational entropy of the particles. However, mechanically, this pressure manifests in different ways depending on the details of the particle dynamics. In the case of colloidal particles \textit{with inertia} $\Pi^B$ arises from the momenta of the particles with $\Pi^B = n \langle m\bm{U \cdot U}/2 \rangle$ where $m$ is the mass of the particle and $\langle ... \rangle$ represents an average over all particles. From equipartition it follows that $\Pi^B = nk_BT$. 

In the case of overdamped dynamics, the particle momentum is ill-defined and the Brownian osmotic pressure must be thought of us a \textit{diffusive pressure}. To see this, we use the standard virial approach for computing the stress (e.g.,~taking the first spatial moment of the Brownian force) and find $\Pi^B = n\langle \bm{x}\cdot\bm{F^B} \rangle = n\zeta \int\langle \bm{U^B}(t')\cdot \bm{U^B}(t)\rangle dt$ where we now recognize  $\int\langle \bm{U^B}(t')\cdot \bm{U^B}(t)\rangle dt$ as the particle's Brownian diffusivity $D^B$ and $\Pi^B = n\zeta D^B$. The Brownian velocity is trivially related to the Brownian force $\bm{U^B} = \bm{F^B}/\zeta$ and is therefore also $\delta$-correlated in time. This has the important implication that the Brownian osmotic pressure can be measured at any instant in time as the diffusivity can be instantaneously measured, in contrast to the swim diffusivity of active matter which requires a duration of $\tau_R$ before it can be measured~\cite{Takatori2014}.

Computing the local pressure of interacting particles with inertia can be readily achieved using the method-of-planes procedure~\cite{Todd1995} which, in a nutshell, computes the sum of interparticle force acting across a surface (see Fig.~\ref{fig:forceflux}) and the rate of change of momentum in the control volume due to particles entering and exiting the surfaces. Importantly, the latter two fluxes can be measured \textit{instantaneously} at the surface. We now propose an extension of the method-of-planes procedure to measure the local stresses generated by single-body forces, such as the Brownian and swim forces. 

As a particle moves across a surface with a Brownian $F^{B}$ and swim force $F^{\text{swim}}$, it will exert a surface force on the particles in the neighboring control volume. We can compute this force by asking a simple question: what is the force required to keep the particles from crossing the imaginary surface plane? In other words, if the imaginary surface was an infinitely thin impenetrable wall, what force would the particles exert on the wall? This is precisely what we compute in Fig.~3 in the main text: at each simulate timestep we compute the force per unit area a hard wall would exert on the particles $\mathcal{F}$. The forces this hypothetical wall must exert to counteract the Brownian $\mathcal{F}^{B}$ and swim forces $\mathcal{F}^{\text{swim}}$ can be readily distinguished (with $\mathcal{F} = \mathcal{F}^{B} + \mathcal{F}^{\text{swim}}$). We note, however, that these quantities are not entirely decoupled as the frequency a particle crosses the surface is a function of both the Brownian and swim force. We further note that, as $\mathcal{F}$ should be interpreted as a local stress, its value is independent of which side of the hypothetical wall is used to measure it.  

Single-body forces only contribute to surface-force flux \textit{the instant} the particle is at the surface. It is for this reason that, when computing the swim force flux, we recover a local swim stress that is vanishingly small in comparison to the magnitude of the swim pressure. The swim diffusivity requires knowledge of the trajectory (the run length) of the particle. In contrast, the Brownian force at any instant in time fully encapsulates the Brownian diffusivity $D_T$ of the particle and results in finding exactly the anticipated diffusive pressure $\mathcal{F}^B = n(z)\zeta D_T = n(z)k_BT$ at any instant in time and, hence, space, as shown in Fig.~3 in the main text. 

\section{Movies}
The file ``dropletdynamics.mp4" contains a visualization of the dynamics of the active liquid droplet shown in Fig.~1A in the main text. The total duration of the video is $\approx 1100 a/U_0$. ``slabpreparation.mp4" illustrates our procedure for creating a single-liquid domain in the slab geometry.